\documentclass[reprint,
superscriptaddress,
nofootinbib,
aps,
pra,
showkeys
]{revtex4-2}
\pdfoutput=1

\usepackage[T1]{fontenc}
\usepackage[utf8]{inputenc}

\usepackage{mathtools, amssymb, mathrsfs, dsfont}
\usepackage{orcidlink}
\usepackage{braket, physics}

\usepackage{graphicx}
\usepackage[export]{adjustbox}
\usepackage{hyperref,xcolor}
\usepackage{placeins}

\definecolor{teal}{HTML}{008080}

\usepackage[nameinlink,capitalize]{cleveref}
\hypersetup{
  colorlinks   = true,
  linkcolor    = teal,
  citecolor    = teal,
  urlcolor     = teal
}
\usepackage{comment}

\newcommand{\Id}{\mathds{1}}
              
       \newcommand{\piZero}{\pi_0(\beta)}             

\newcommand{\JJ}[1]{\mathcal{J}_{#1}}          
\newcommand{\JJi}[1]{\mathcal{J}^{-1}_{#1}}    

\newcommand{\SLD}{L}                           \newcommand{\FQ}{\mathcal{F}}                  
\DeclareMathOperator{\Var}{Var}

\newcommand{\Ut}{U_{t,\beta}}
\newcommand{\Utd}{U_{t,\beta}^\dagger}

\newcommand{\lamt}{\lambda(t,\beta)}
\newcommand{\SLDof}[1]{\SLD_{#1}^\beta}      
\newcommand{\FQof}[1]{\FQ^\beta_{#1}}  
\newcommand{\FQofSTAR}[1]{\FQ^{\beta^*}_{#1}} 
\newcommand{\FQt}{\FQ^\beta_t}                
             
\newcommand{\It}{\mathcal{I}^\beta_t}       

\newcommand{\teal}[1]{\textcolor{black}{#1}}

\begin{document}

\title{Shake before use: universal enhancement of quantum thermometry by unitary driving}
	\author{Emanuele Tumbiolo\,\orcidlink{0009-0000-6411-1285}}
    \email[Corresponding author: ]{emanuele.tumbiolo01@ateneopv.it}
    \affiliation{Dipartimento di Fisica, Università degli Studi di Pavia, Via Agostino Bassi 6, I-27100, Pavia, Italy}
    \affiliation{INFN Sezione di Pavia, Via Agostino Bassi 6, I-27100, Pavia, Italy}
	
\author{Lorenzo Maccone\,\orcidlink{0000-0002-6729-5312}}
    \affiliation{Dipartimento di Fisica, Università degli Studi di Pavia, Via Agostino Bassi 6, I-27100, Pavia, Italy}
    \affiliation{INFN Sezione di Pavia, Via Agostino Bassi 6, I-27100, Pavia, Italy}
    
\author{Chiara Macchiavello\,\orcidlink{0000-0002-2955-8759}}
    \affiliation{Dipartimento di Fisica, Università degli Studi di Pavia, Via Agostino Bassi 6, I-27100, Pavia, Italy}
    \affiliation{INFN Sezione di Pavia, Via Agostino Bassi 6, I-27100, Pavia, Italy}

\author{Matteo G. A. Paris\,\orcidlink{0000-0001-7523-7289}}
    \affiliation{Dipartimento di Fisica, Universit\`a di Milano, I- 20133, Milano, Italy}

\author{Giacomo Guarnieri\,\orcidlink{0000-0002-4270-3738}}
    \affiliation{Dipartimento di Fisica, Università degli Studi di Pavia, Via Agostino Bassi 6, I-27100, Pavia, Italy}
    \affiliation{INFN Sezione di Pavia, Via Agostino Bassi 6, I-27100, Pavia, Italy}

\begin{abstract}
Quantum thermometry aims at determining temperature with ultimate precision in the quantum regime. Standard equilibrium approaches, limited by the Quantum Fisher Information given by static energy fluctuations, lose sensitivity outside a fixed temperature window. Non-equilibrium strategies have therefore been recently proposed to overcome these limits, but their advantages are typically model-dependent or tailored for a specific purpose. This Letter establishes a general, model-independent result showing that any temperature-dependent unitary driving applied to a thermalized probe enhances its quantum Fisher information with respect to its equilibrium value. Such information gain is expressed analytically through a positive semi-definite kernel of information currents that quantify the flow of statistical distinguishability. 
Our results, \teal{together with an analysis of the relation between information gain and control cost}, are benchmarked on a driven spin-$1/2$ thermometer, furthermore showing that resonant modulations remarkably restore the quadratic-in-time scaling of the Fisher information and allow to shift the sensitivity peak across arbitrary temperature ranges.
\end{abstract}
\maketitle
\paragraph*{Introduction. -- }

Unlike standard observables in quantum mechanics, temperature is notoriously not associated with a Hermitian operator and must be instead indirectly inferred~\cite{book:Gemmer_QuantThermo,art:Puglisi_TemperatureRev,art:Mehboudi_ThermometryReview,thes:Jorgensen}. 
In equilibrium quantum thermometry, a probe, hereby referred to as thermometer, is allowed to thermalize with a bath, and the temperature is extracted from the statistics of energy measurements. 
According to quantum-estimation theory~\cite{art:Helstrom_EstimationTheory,art:Giovannetti_Beating,art:Giovannetti_QMetro}, the ultimate precision for any temperature estimation protocol is fixed by the quantum Fisher information (QFI) through the Cramér–Rao bound~\cite{art:Paris_MetrologyReview,art:Braunstein_GeoOfQStates}. For thermal states the QFI equals the thermometer's energy variance, and is thus proportional to its heat capacity~\cite{art:Mehboudi_ThermometryReview,art:Correa_OptimalProbes}. 
Yet equilibrium thermometry is fundamentally limited.
Because Gibbs states commute with their Hamiltonian, the QFI coincides with the classical Fisher information of the thermal population distribution, with no contributions from coherence. Moreover, as one reaches equilibrium, the QFI saturates
and does not increase with the interrogation time. Last but not least, high sensitivity can be crucially achieved only within a fixed and non-tunable temperature window, dictated by the probe energy spectrum, and vanishes both in the low- and high-temperature limits~\cite{art:Potts_FundamentalLimits,art:Jorgensen_Tight}.
In practice, this hinders experimental control over the operating range of any equilibrium thermometer.\\
These limitations have motivated the development of alternative, non-equilibrium approaches, such as interferometric~\cite{chap:DePasquale_QuantumThermo,art:Jarzyna_Interferometric,art:Yang_IP,art:Akamatsu_NonLinMZ} and control-assisted~\cite{art:Mehboudi_FLimitsAdaptiveStrats,art:Kiilerich_AAssisted,art:Sekatski_Optimal} schemes, where external drivings~\cite{art:Mukherjee_Driving,art:Glatthard_BendingRules} or transient dynamics~\cite{art:Razavian_SingleQubit,art:Gebbia_TwoQubits,art:Zhang_SeqMeas,art:Yang_Sequential} are exploited to enhance temperature sensitivity. 
Despite substantial progress, existing results remain largely model-dependent, and a unifying principle quantifying the fundamental role of unitary driving in quantum thermometry has insofar remained elusive.
Here we establish a model-independent result: any unitary perturbation applied to a probe initially in thermal equilibrium cannot degrade its temperature sensitivity, and in fact enhances it under mild and physically reasonable assumptions.
In particular, our main result is an analytical expression for the time-dependent QFI for a generic unitary driving, in the decomposed form
\begin{equation}
    \label{eq:main}
    \FQt = \FQof{\pi_0}+ \It, 
    \qquad 
    \It \ge 0,
\end{equation}
where $\FQof{\pi_0}$ is the equilibrium value and $\It$ quantifies the non-negative contribution arising from the unitary driving. 
A general conclusion following from \cref{eq:main} is that \teal{temperature-dependent unitary drivings implemented on thermalized probes also enable the shift-and-reshape} of the QFI profile across temperatures, effectively relocating the thermometer’s high-sensitivity region to any desired range.
This result holds for arbitrary finite-dimensional probes initially in a full-rank \teal{equilibrium} state. \teal{Extensions to more general regimes (e.g., infinite-dimensional systems, rank-deficient or non-equilibrium initial states), although possible, lie beyond the scope of this work.}
Although exemplified in the context of thermometry, our result extends directly to quantum estimation theory at large, providing a benchmark for precision enhancement in driven quantum systems.

\begin{figure}[htbp]
    \centering
    \includegraphics[valign = c, width = 0.45\textwidth]{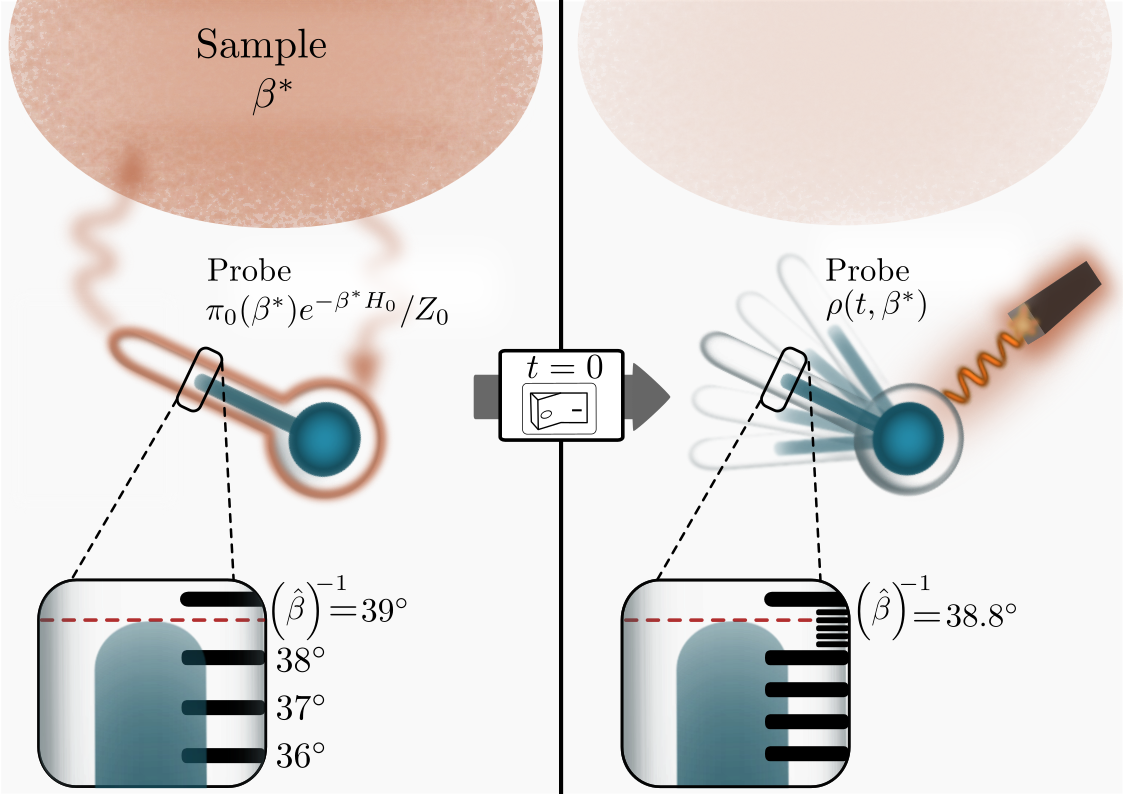}
    \caption{Schematic illustration of the protocol. In the initial stage (left), the probe is allowed to thermalize with the sample, encoding the unknown inverse temperature $\beta^*$ into the Gibbs state $\pi_0(\beta^*)$, and yielding an estimation precision bounded by the equilibrium quantum Cramér–Rao bound in \cref{eq:CramerRaoB}. At $t=0$ the driving is switched on, perturbing the now isolated probe. The resulting QFI of the non-equilibrium state $\rho(t,\beta^*)$ evolves according to \cref{eq:main}, increasing whenever the drive depends non-trivially on temperature and thereby enhancing the attainable precision in the estimation (here illustrated as the acquisition of an additional significant digit in the estimator $\hat{\beta}$).}
    \label{fig:Schematics}
\end{figure}

\paragraph*{Equilibrium thermometry. --}
In order to determine the temperature $(\beta^*)^{-1}$ (in units where $k_B = 1$) of a target sample, acting as a thermal bath, the minimal strategy is to infer it from a probe (i.e. the thermometer) that has interacted with it until thermalization has occurred. 
Owing to the so-called ``zero-th law"~\cite{book:Planck_Thermo, book:Callen_Thermo} of Thermodynamics, such thermalization process in fact generally encodes the temperature $\beta$ of a generic sample into the probe, leading to the definition of a smooth one-parameter statistical manifold $\{\piZero\}_\beta$, with $\piZero = e^{-\beta H_0}/Z_0$ being the Gibbs state of the probe with respect to its bare Hamiltonian $H_0$. 
Within local quantum estimation theory, the maximum achievable precision in inferring the true value of the parameter $\beta = \beta^*$ for any unbiased estimator $\hat{\beta}$ based on $n$ independent measurements is set by the quantum Cramér-Rao bound
\begin{equation}
  \label{eq:CramerRaoB}
  \Var[\hat{\beta}] \ge \frac{1}{n\,\FQofSTAR{\pi_0}}\,\,,
\end{equation}
where $\FQofSTAR{\pi_0} = \FQof{\pi_0}|_{\beta=\beta^*}$ is the QFI, which represents the maximum of the classical Fisher information optimized over all positive operator-valued measures, and quantifies the infinitesimal distinguishability between states in the statistical model.
For an arbitrary smooth family of states $\sigma(\beta)$, the QFI can be explicitly expressed as 
\begin{align}
\label{eq:FisherDef}
\FQof{\sigma} = \Tr\left[(\partial_\beta \sigma)\SLDof{\sigma}\right]
= \Tr\left[\sigma\,(\SLDof{\sigma})^2\right],
\end{align}
where the symmetric logarithmic derivative (SLD)~\cite{art:Paris_MetrologyReview} $L_\sigma^\beta$ is the Hermitian operator satisfying the Lyapunov-type equation
\begin{equation}
\label{eq:SLDDef}
\partial_\beta \sigma = \JJ{\sigma}(L_\sigma^\beta), 
\qquad
\JJ{\sigma}(X) := \tfrac{1}{2}\{\sigma,X\}.
\end{equation}
The Bures-Jordan multiplication superoperator $\JJ{\sigma}$~\cite{art:Scandi_WorkStats,art:Abiuso_FLimitsMetro} induces a metric on the quantum statistical manifold parametrized by $\beta$ through the inner product
\begin{equation}
\langle\!\langle A,B \rangle\!\rangle _{\sigma} = \,\Tr[A^\dagger\,\JJ{\sigma}(B)].
\end{equation}
Accordingly, the SLD $L_\sigma^\beta$ is the metric dual~\cite{art:Ciaglia_DiffGeoOfStates, book:Petz_QI,book:Amari_InfoGeometry} of the tangent vector $\partial_\beta\sigma$, i.e. 
$L_\sigma^\beta = \JJi{\sigma}(\partial_\beta \sigma)$, 
where, for full-rank $\sigma$, an explicit representation of the inverse Bures superoperator is
$\JJi{\sigma}(X)=2\int_0^\infty ds\, e^{-s\sigma}\, X \,e^{-s\sigma}$.
At thermal equilibrium, $[\pi_0,H_0]=0$ implies
\begin{align}
  \label{eq:EquilibriumSLD}
  L_{\pi_0}^\beta &= -\big(H_0 - \Tr[H_0\,\pi_0]\big),\\
  \label{eq:EquilibriumFisher}
  \FQof{\pi_0} &= \Var_{\pi_0}[H_0],
\end{align}
which is proportional to the heat capacity of the probe up to a $\beta^2$ factor~\cite{art:Paris_LandauLimit}. 
Because \cref{eq:CramerRaoB}, with $\FQofSTAR{\pi_0} = \left.\FQof{\pi_0}\right|_{\beta=\beta^*}$, is saturated by energy measurements, equilibrium thermometry is effectively classical and the achievable precision is fixed by static energy fluctuations. 
As the probe approaches $\pi_0$, the QFI asymptotically saturates to \cref{eq:EquilibriumFisher}, independently of the interrogation time $t$, in contrast with unitary metrological protocols where phase accumulation yields the characteristic quadratic scaling $\FQof{\sigma} \propto t^2$~\cite{art:Das_UniversalTimeScal}. 
Moreover, the sensitivity window of equilibrium thermometry is static and determined by the spectrum of $H_0$: for gapped systems the QFI is exponentially suppressed in both the $\beta\to 0$ (maximally mixed) and $\beta\to\infty$ (ground-state) limits, whereas for gapless ones it decays algebraically~\cite{art:Paris_LandauLimit,art:Mehboudi_ThermometryReview}. 
These intrinsic limitations have motivated the exploration of non-equilibrium and control-assisted thermometric strategies, in which external drivings, engineered couplings~\cite{art:Planella_BathInduced,art:Abiuso_SpinNetThermometer,art:Kiilerich_AAssisted}, or memory effects~\cite{art:Aiache_NonMarkNonEq,art:Zhang_SeqMeas,art:Yang_Sequential} are exploited to activate coherence and modify the scaling and range of the QFI.
A paradigmatic instance is that of periodically driven thermometers, where modulation of the probe spectrum can alter the low-temperature scaling of the achievable precision~\cite{art:Glatthard_BendingRules}. 
While such schemes demonstrate that control can outperform static thermometry, they remain tied to specific dynamics and platforms: a general, model-independent understanding of how driving can reshape the fundamental bounds on thermal sensitivity is still lacking.
\begin{figure*}[htpb]
    \centering
    \includegraphics[valign = c, width = 0.9\textwidth]{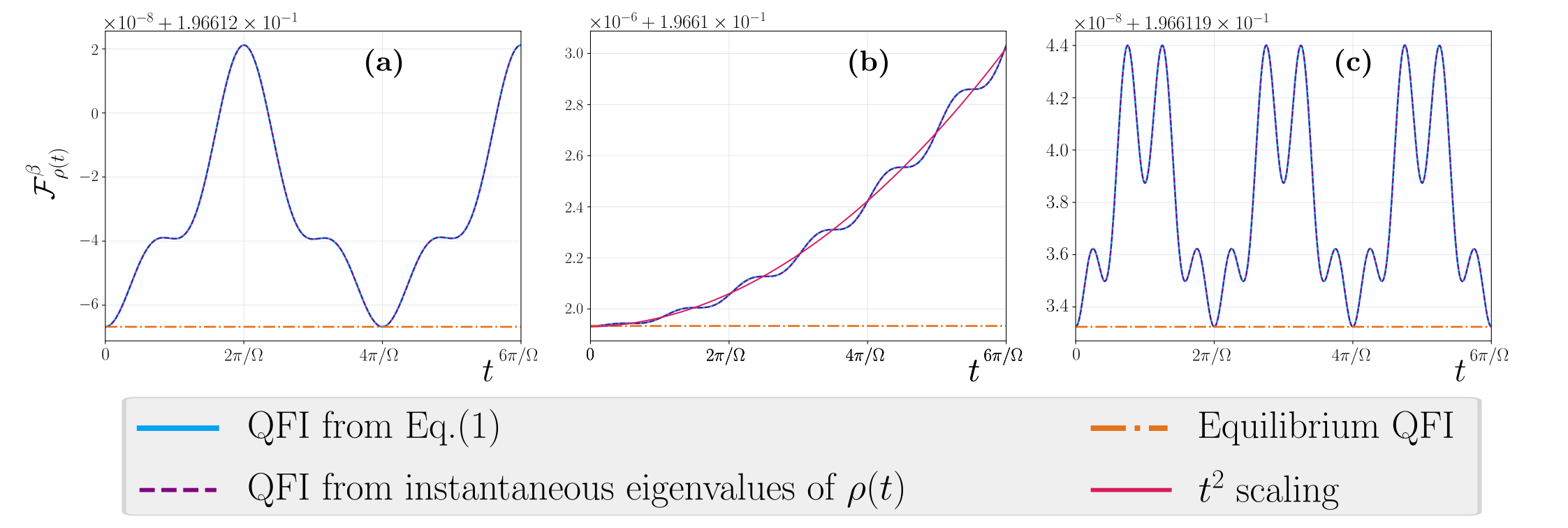}
    \caption{Time evolution of the QFI for a single-spin probe under a 
    temperature-dependent transverse driving. Time $t$ is measured in units of the spin characteristic 
    period $2\pi/\Omega$, with $\Omega=1$.
    Solid and dashed lines denote, respectively, the QFI as given by \cref{eq:main} and by the standard
    expression in terms of the instantaneous eigenvalues of $\rho(t)$~\cite{art:Liu_QFIM}, showing perfect agreement. The 
    dash-dotted line marks the equilibrium QFI benchmark. 
    The three panels illustrate the transition from bounded dynamics 
    to resonant quadratic scaling, identifying the optimal driving frequency. 
    The Gaussian temperature control profile is chosen with a standard deviation saturating 
    the equilibrium quantum Cramér–Rao bound, as if an optimal equilibrium estimation 
    had been performed beforehand, and is randomly centered within the interval $\max\bigl(0,\,\beta^* \pm 1/\sqrt{\FQofSTAR{\pi_0}}\bigr)$. For all panels $\phi = 0 $ and $\lambda_0 = 0.1$. 
    \textbf{(a)} Off-resonant regime with $\omega_{\mathrm d}=0.5\,\Omega$, 
    where the QFI exhibits bounded oscillations.
    \textbf{(b)} Resonant regime with $\omega_{\mathrm d}=\Omega$, where the 
    QFI displays an almost monotonic $t^2$ growth, 
    signalling a build-up of sensitivity. 
    \textbf{(c)} High-frequency regime with $\omega_{\mathrm d}=2\,\Omega$, where one observes once again a bounded, oscillatory response, albeit at a faster characteristic timescale.}
    \label{fig:FrequencyScan}
\end{figure*}
\paragraph*{Driving a thermometer. --}
To single out the role of unitary control in determining temperature sensitivity, we consider the minimal scenario in which the probe, having first thermalized with the sample, is then subjected to an external perturbation mediated by a classical field, as depicted in \cref{fig:Schematics}. \teal{We stress that, at this stage, any interaction between probe and sample has been switched off: the probe is fully isolated and the sample is excluded from the subsequent dynamics.}
In this regime the dynamics is unitary, consistently with the standard formulation of quantum optimal control theory.
The evolution is described by the propagator $\Ut$, generated by the time-dependent Hamiltonian
\begin{equation}
\label{eq:Driving}
  H(t,\beta)=H_0+\lamt V,
\end{equation}
where  in general $[H_0,V]\neq0$ and $\lamt$ is the driving profile, smooth in $\beta$.
The explicit temperature dependence of the drive encompasses the widest class of perturbations one can consider.
From a thermodynamical perspective, the control field $\lambda(t,\beta)$ \teal{may arise} within an open quantum systems framework when the probe is coupled to an auxiliary work-reservoir, itself in equilibrium at the same temperature as the sample~\cite{art:Colla_WorkReservoirs,book:Strasberg_StochasticThermo} \teal{(for further details, see Section 2 of~\cite{ref:SM})}\nocite{art:Petz_GeometryQuStates,art:Morozova_Geometry,thes:Scandi_PhD,book:Balian,art:Bottcher, art:Wenzel, art:Cheng,art:Kittaneh,art:WerschnikQOCT}. In this regime, such a reservoir behaves in fact as an ideal work source with negligibly small fluctuations, unitarily exchanging energy with the probe and thus providing a deterministic, non-dissipative drive: tracing out its degrees of freedom yields an effective Hamiltonian correction on the probe whose amplitude inherits its $\beta$-dependence directly from the reservoir’s thermal state. 
Operationally, one may \teal{tune} the driving profile in \cref{eq:Driving} adaptively, consistently with a local estimation framework: first of all, a preliminary coarse-grained measurement of the equilibrium probe provides a prior estimate $\beta_0\approx \beta^*$ via \cref{eq:EquilibriumFisher},
which is then used as a working point~\cite{art:Giovannetti_Advances} \teal{to shift the operating region} of the control field and \teal{maximize} the sensitivity in a neighborhood of $\beta_0$. In the following, $\beta_0, s_\beta$ are to be interpreted as fixed design parameters.
\teal{For concreteness,} in the rest of the paper we consider drivings of the form
\begin{equation}
  \label{eq:GenericLambda}
  \lamt=\lambda_0 \, G_{\beta_0,s_\beta}(\beta) \, f(t),
\end{equation}
where $\lambda_0$ is the driving strength, $f(t)$ its temporal modulation, and 
$G_{\beta_0,s_\beta}(\beta) = \exp[-(\beta-\beta_0)^2/(2s_\beta^2)] \equiv G(\beta)$ a dimensionless Gaussian envelope centered around the initial estimate $\beta_0$, with standard deviation 
$s_\beta$. \teal{We remark that the above choice \cref{eq:GenericLambda} is purely illustrative (one could consider any physically motivated smooth temperature dependence) and does not affect the generality of the subsequent results.}
To assess how such drivings modify thermal sensitivity, we analyse the corresponding family of evolved states 
 $\rho(t) = \Ut \, \piZero \,\Utd $.
Exploiting the unitary covariance of $\JJ{\sigma}$ and its inverse, the corresponding SLD can be expressed as
\begin{equation}
  \SLDof{\rho} = \Ut\,\big(L_{\pi_0}^\beta+\delta L_t^\beta\big)\,\Utd,
\end{equation}
with $L_{\pi_0}^\beta$ as in \cref{eq:EquilibriumSLD}, while
\begin{equation}
  \delta L_t^\beta = \int_0^t ds\,\partial_\beta\lambda(s)\,J_V(s)
\end{equation}
captures the non-equilibrium contribution induced by the temperature dependence of the driving. 
The operator  
\begin{equation}
\label{eq:JVdef}
J_V(s) = -\frac{i}{\hbar}\,\JJi{\pi_0}\!\left([V_H(s),\pi_0]\right),
\end{equation}
here introduced, which we call information current, quantifies the instantaneous flow of statistical distinguishability generated by the perturbation $V_H(s)$ (in the Heisenberg picture) with respect to the equilibrium reference state $\pi_0$.  
It vanishes whenever $[V_H(s),\pi_0]=0$, i.e. when no quantum friction is generated through the driving, since the latter leaves the eigenbasis of $\pi_0$ invariant, acting only on its eigenvalues~\cite{art:Feldmann_Friction,art:Plastina_Friction,art:Miller_Friction,Onishchenko_TrappedIonFriction,art:Francica_Friction}.
Hence, a nonzero $J_V(s)$ signals that the control field injects coherences in the eigenbasis of $\pi_0$: the ensuing enhancement of distinguishability thus stems from a genuinely quantum mechanism,  \teal{as further discussed in Section 4 of ~\cite{ref:SM}}.
Similarly, when $\partial_\beta\lambda=0$, i.e. when the driving is not sensitive to temperature, one recovers $L_{\rho(t)}^\beta=L_{\pi_0}^\beta$.
This identifies a genuine no-go condition, as it excludes any unitary control that does not explicitly depend on $\beta$ from generating additional information about temperature, regardless of its time dependence or strength.
Geometrically, this reflects the fact that a parameter-independent unitary is an isometry of the Bures metric, rotating the probe state within the statistical manifold without changing its sensitivity to temperature. This conclusion is fully consistent with the quantum data-processing inequality~\cite{art:Petz_Monotones}, which forbids any Fisher information gain under parameter-independent completely positive trace-preserving maps, including unitaries.

Using \cref{eq:FisherDef} and noticing that $\Tr[\pi_0\,\delta L_t^\beta\,L_{\pi_0}^\beta]=0$, allows to derive~\cite{ref:SM} our main result \cref{eq:main}, which quantifies the QFI increment due to a generic unitary driving with
\begin{equation}
\label{eq:kernel}
\begin{aligned}
  \It
  &=\Tr\left[\piZero\left(\delta L_t^\beta\right)^2\right] = \\
  &= \int_0^t \int_0^t
  \partial_\beta\lambda(s,\beta)\,\partial_\beta\lambda(u,\beta)\,
  K(s,u,\beta)\,ds\,du\,
  .
\end{aligned}
\end{equation}
which is manifestly non-negative, and where we have introduced the kernel
\begin{equation}
  K(s,u,\beta)=\Tr[\piZero\,J_V(s)J_V(u)]
\end{equation}
which encodes the two-time correlation function of the information current. 
We remark once again that the QFI must be evaluated at $\beta=\beta^*$ and that, if the perturbation commutes with the initial Hamiltonian, i.e. $\left[ H_0, V\right]=0$, then the improvement vanishes and no precision enhancement is expected.

Interestingly, by interpreting the external driving as part of a measurement scheme, our results can be recast within the framework of quantum parameter estimation with parameter-dependent measurement apparatuses, put forward in Ref.~\cite{art:Seveso_Beyond}, where the standard Cramér-Rao bound may in principle be surpassed. In the present work, however, we focus explicitly on the modification of the QFI induced by the $\beta$-dependent dynamics, without optimizing the drive. In this sense,  \cref{eq:main} holds true for any arbitrary choice of driving provided it depends non-trivially on temperature.

The positivity of \cref{eq:kernel} thus provides a measure of how temperature-dependent drivings amplify thermal sensitivity.
At short times, regularity of the kernel guarantees that $\It \sim t^2$, reproducing the standard scaling of the QFI in closed evolutions where distinguishability grows linearly in time at the level of the SLD. 
Importantly, the same quadratic scaling can re-emerge asymptotically when the driving preserves constructive correlations in the information current, such as for constant (in time) controls or for periodic modulations resonant with the Bohr frequencies of $H_0$. This guarantees the scaling of the QFI with $t$ to be quantum-like, i.e. quadratic, at all times~\cite{art:Das_UniversalTimeScal}.
Since the kernel depends on the $\beta$-derivative of the control profile $\lamt$, shaping this dependence provides a direct way of shifting the sensitivity window of the probe. 
In this sense, the driving acts as an operational handle on the probe sensitivity landscape, and can naturally accommodate feedback-assisted strategies in which $\lamt$ is updated according to intermediate estimates of $\beta$, keeping the probe dynamically locked to the region of maximal information gain.\\
To benchmark our results, we analyze a single spin-$1/2$ probe as a minimal reference model in which all relevant mechanisms can be made explicit.

\paragraph*{Single spin probe. --}
In this minimal setting, the dynamics is governed by a driven two-level Hamiltonian,
\begin{equation}
H(t,\beta)=\frac{\Omega}{2}\sigma_z+\lamt\sigma_x,
\end{equation}
where $\Omega$(in units of $\hbar=1$)  denotes the characteristic frequency of the system, while $\lamt$ is chosen 
according to \cref{eq:GenericLambda}, with periodic temporal modulation
$f(t)=\cos(\omega_{\mathrm d}t+\phi)$, driving frequency $\omega_{\mathrm d}$ and phase $\phi$.
In the absence of control, the probe is allowed to reach thermal equilibrium, characterized by the baseline Fisher information
\begin{equation}
\FQof{\pi_0}
=\Big(\tfrac{\Omega}{2}\Big)^{\!2}
\!\mathrm{sech}^2\!\Big(\tfrac{\beta\Omega}{2}\Big).
\end{equation}
Switching on the drive turns on the information current, leading to the emergence of a non-trivial $K(s,u,\beta)$. 
\begin{figure}[h!]
    \centering
    \includegraphics[valign = c, width = 0.45\textwidth]{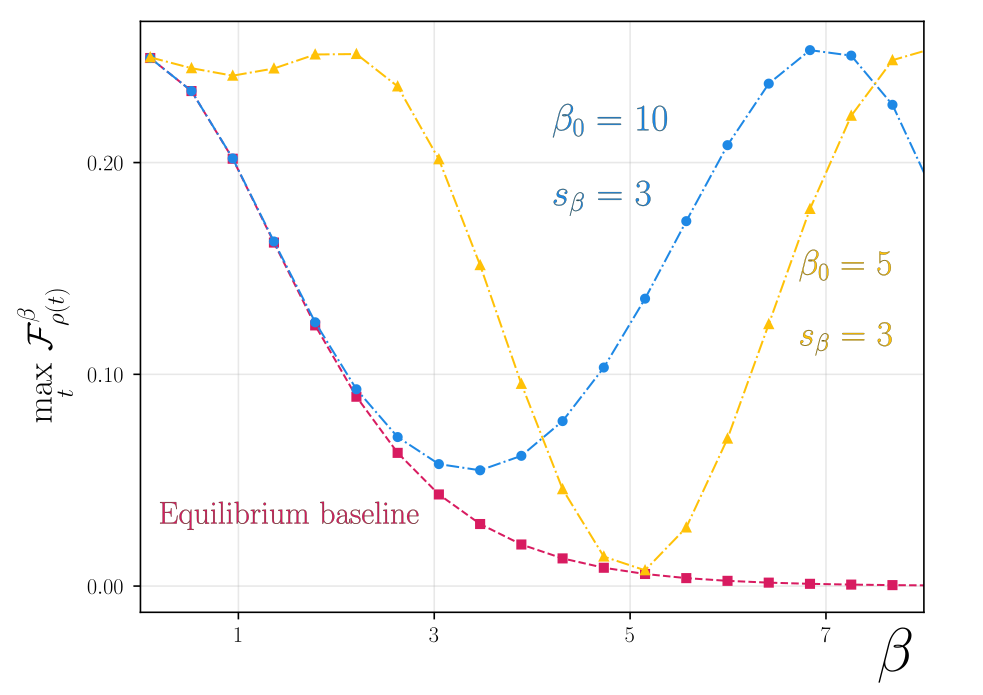}
     \caption{
    Behaviour of the maximum QFI at a fixed evolution time $t=12\,s$ in the resonant regime, as a function of the inverse temperature $\beta$.
    The magenta dashed line (with squares) represents the equilibrium baseline
    $\FQof{\pi_0}$, while the blue circles and yellow triangles show the maxima for the driven case,
    for $G(\beta)$ with $(\beta_0,s_\beta)=(10,3)$ and $(5,3)$, respectively.
    By tuning the mean and standard deviation of the Gaussian envelope, the sensitivity window can be
    shifted to different temperature ranges.
    The resulting QFI profiles display two symmetric lobes around
    $\beta_0$, reflecting the underlying Gaussian modulation of the driving
    amplitude, as exemplified by the yellow curve.}
    \label{fig:ShiftingSensitivity}
\end{figure}

The time integral governing $\It$ contains harmonic components oscillating at $\Omega\pm \omega_d$ , with the detuning $\Omega-\omega_d$ determining whether correlations in the information current interfere constructively or destructively over time.
\cref{fig:FrequencyScan} illustrates this effect  for three representative driving frequencies: below, on, and above resonance.
On resonance, i.e. $\omega_{\mathrm d}\simeq\Omega$, and in a weak-field approximation, the increment $\It$ can be shown explicitly (See Section 3 of~\cite{ref:SM}) to achieve  quadratic scaling for 
long times
\begin{equation}
\It \approx
\big(\lambda_0 G'(\beta)\big)^2
\alpha(\beta)\,t^2\,,
\label{eq:LongT_MainText}
\end{equation} with $G'(\beta) = \partial_\beta G(\beta) $ and $\alpha(\beta) $ being a time-independent quantity dictated only by the equilibrium state of the thermometer, 
confirming that resonance restores the fully unitary $t^2$ scaling discussed above.  
For detuned drivings ($\omega_{\mathrm d}\neq\Omega$), instead, the oscillatory kernel induces dephasing that suppresses the long-time growth and saturates the information gain.
In \cref{fig:ShiftingSensitivity}, we fix $\omega_{\mathrm d} \simeq \Omega $ and vary the center and standard deviation of the Gaussian temperature envelope, showing how the QFI can be reshaped and shifted across distinct temperature regions.
This demonstrates that the probe’s sensitivity window can be tailored by tuning the functional dependence of the control amplitude on $\beta$.

\teal{Finally, we investigate the energetic cost associated with the external driving that is responsible for the predicted information gain.
To this end, inspired by shortcuts-to-adiabaticity and counterdiabatic-driving methods~\cite{art:ZhengCostCD,art:CampbellTradeoffSTA}, we introduce the following cost functional to quantify the accumulated driving effort
\begin{equation}
\mathcal{C}_2(t,\beta):=\int_0^t ds\,|\lambda(s,\beta)|^2\, \|V\|_2^2 ,
\end{equation}
with $\|V\|_2$ denoting the Frobenius norm. This allows us to further consider a dimensionless ratio measuring the information gain per unit of energetic cost
\begin{equation}\label{eq:costratio}
\mathcal{R}_2(t,\beta):=\frac{\It}{\mathcal{C}_2(t,\beta)} \;.
\end{equation}
In the time-asymptotic limit of our model, we find that $\mathcal{R}_2(t,\beta)\sim t$ in the resonant regime, while it vanishes away from resonance (see Section 5 of ~\cite{ref:SM} for further details). 
This clearly emphasizes the additional efficacy of driving the spin-$1/2$ probe on resonance, since a linearly growing energetic cost $\mathcal{C}_2(t,\beta)\sim t$ would allow for a quadratically increasing information gain $\It \sim t^2$.}

\paragraph*{Conclusion. --}
In this work we have derived a general analytic expression for the temperature sensitivity, i.e. its Quantum Fisher Information, ensuing from any temperature-dependent unitary driving acting on an initially thermalized probe. 
Our results show that implementing such a perturbation increases temperature precision, quantified by a non-negative correction to the equilibrium heat capacity of the thermometer expressed in terms of a kernel of information-geometric currents that captures the unitary build-up of temperature information along the drive. We then showcase our findings on a driven spin-$1/2$ model, remarkably recovering a $t^2$ scaling of the QFI in the resonant-driving regime. 

Furthermore, we explicitly demonstrate that the sensitivity window of the qubit probe can be tuned across the whole temperature range by suitably acting on the driving parameters.
We want to emphasize that the framework and the results here explicitly obtained within the context of quantum thermometry can be straightforwardly applied to study the enhancement in the estimation of other parameters.
Finally, it is worth mentioning that the predicted sensitivity increase and tunability can be directly applicable to current experimental setups for quantum thermometry such as bichromatically driven trapped ions~\cite{art:Li_TrappedIons} or microwave-driven dispersive circuit QED~\cite{art:Gambetta_MeasIndDephasing,art:Sears_DispersiveCQED,art:Atalaya_MeasSmallPhotonNums}.
A natural next step will be to allow for the control field to act already during thermalization, or to consider non-unitary drivings; \teal{in this case, the noise introduced by dissipative channels may in general compete with the information gain, reducing the maximum achievable sensitivity~\cite{art:Petz_Monotones,art:Escher}. 
Our results therefore pave the way for future studies characterizing the general trade-off between unitary control and dissipation in thermal sensitivity enhancements.}

\section*{Acknowledgments}
E.T. acknowledges support from the PNRR MUR Project PE0000023-NQSTI. L.M. acknowledges support from the PRIN MUR Project 2022RATBS4.
C.M. acknowledges support from the National Research Centre for HPC, Big Data and Quantum Computing, PNRR MUR Project CN0000013-ICSC.  M.G.A.P acknowledges  partial support by Khalifa University of Science and Technology through the project C2PS-8474000137. G.G. kindly acknowledges support from the Ministero dell’Università e della Ricerca (MUR) under the “Rita Levi-Montalcini” grant and to INFN.
\bibliography{refs.bib}
\pagebreak

\widetext
\begin{center}
\vskip0.5cm
{\Large \textbf{Supplemental Material}}
\end{center}
\vskip0.4cm

\setcounter{section}{0}
\setcounter{equation}{0}
\setcounter{figure}{0}
\setcounter{table}{0}
\setcounter{page}{1}
\renewcommand{\theequation}{S\arabic{equation}}
\renewcommand{\thefigure}{S\arabic{figure}}

\section*{}\teal{This Supplemental Material provides the detailed derivations and complementary analyses supporting the results of the main text. Section~S1 derives the general decomposition of the quantum Fisher information into an equilibrium contribution and a dynamical increment, and recovers the exact kernel representation governing the latter for arbitrary $\beta$-dependent unitary dynamics. Section~S2 addresses the physical origin of the temperature-dependent driving considered in the main text, identifying concrete microscopic mechanisms through which such a dependence can arise and clarifying its operational interpretation within a local estimation framework, further discussing robustness to calibration mismatches. Extensions beyond strictly unitary dynamics are also briefly mentioned. Section~S3 specializes the general formalism to the driven single-spin probe, providing explicit expressions for the kernel and deriving the corresponding short- and long-time scaling behaviors. Section~S4 further investigates the role of coherences (in the eigenbasis of the equilibrium state) in the generation of information currents, shedding light on how non-commutativity between the initial state and the perturbation constrains the achievable sensitivity enhancement. Finally, Sec.~S5 further analyzes the figure of merit introduced to quantify control cost in the single spin-$1/2$ probe, providing a more detailed discussion and numerical evaluation of its behavior, and assessing the resulting trade-off between metrological gain and control resources.}
\section*{S1: Deriving the expression for the dynamical increment of the QFI}
\label{app:S1}

In this section we derive the two central identities of this work, namely Eqs.~(1) and (13) of the main text. The former expresses the quantum Fisher information of the driven probe at time $t$ as
\begin{equation}
  \FQof{\rho(t)}=\FQof{\pi_0}+\It,
\end{equation}
i.e. as the sum of an equilibrium baseline, fixed by the initial thermal state, and a purely dynamical, non-negative increment $\It$ generated by the drive. Equation~(13), instead, provides an exact closed form for such an increment. 

To set the stage, we recall the standard geometric interpretation of the QFI as a quantifier of local sensitivity in quantum parameter estimation. Since we are concerned with thermometry, we restrict to the one-parameter scenario of the inverse temperature $\beta$, but all considerations extend to the multiparameter case. 

Consider a finite-dimensional thermometer with Hilbert space $\mathcal H$ and a smooth, one-parameter family of full-rank density operators $\{\sigma(\beta)\}$. The mapping $\beta\mapsto\sigma(\beta)$ defines a statistical manifold whose tangent space at each point $\sigma$ consists of traceless Hermitian operators and is spanned by the derivative vector $\partial_\beta\sigma$.

Such a manifold is Riemannian upon the choice of a suitable metric $g_\sigma(X,Y)$ for $X,Y$ in the tangent space at $\sigma$. According to the Morozova–Čencov–Petz classification, the admissible metrics form a continuous family generated by operator-monotone functions. We refer the reader to~\cite{art:Petz_Monotones,art:Petz_GeometryQuStates,art:Morozova_Geometry,thes:Scandi_PhD} for further information.  
In quantum parameter estimation, the canonical and operationally relevant choice is
the Bures metric, induced by the Bures inner product
\begin{equation}
  \langle\!\langle X,Y\rangle\!\rangle_\sigma
  = \Tr\!\big[X^\dagger\,\JJ{\sigma}(Y)\big],
  \label{eq:Bures_inner}
\end{equation}
where $\JJ{\sigma}(Y)=\tfrac{1}{2}\{\sigma,Y\}$ is the Bures–Jordan superoperator with inverse  
\begin{equation}
  \JJi{\sigma}(X)
  = 2\!\int_0^{\infty} e^{-s\sigma}\,X\,e^{-s\sigma}\,ds.
\end{equation}
Both are unitarily covariant:
for an arbitrary unitary $U$,
\begin{equation}
  \JJ{U\sigma U^\dagger}(X)
  = U\,\JJ{\sigma}(U^\dagger XU)\,U^\dagger,
  \qquad
  \JJi{U\sigma U^\dagger}(X)
  = U\,\JJi{\sigma}(U^\dagger XU)\,U^\dagger,
  \label{eq:Bures_covariance}
\end{equation}
a property that will be needed further in the derivation. 
Within this structure
\begin{equation}
  g^{(\mathrm B)}_\sigma(\partial_\beta\sigma,\partial_\beta\sigma)
  = \langle \! \langle \JJi{\sigma}(\partial_\beta\sigma),\JJi{\sigma}(\partial_\beta\sigma)\rangle\!\rangle_\sigma =\Tr\!\big[\partial_\beta\sigma\,\JJi{\sigma}(\partial_\beta\sigma)\big].
  \label{eq:Bures_metric}
\end{equation}
By identifying the dual under this metric of the tangent vector $\partial_\beta\sigma$ as the symmetric logarithmic derivative (SLD),
\begin{equation}
  \SLDof{\sigma}
  := \JJi{\sigma}(\partial_\beta\sigma),
\end{equation}
i.e. the unique solution to the Lyapunov-type equation $\partial_\beta\sigma
  = \tfrac{1}{2}\{\sigma,\SLDof{\sigma}\}$, it is immediate to show that   
\begin{equation}
  g^{(\mathrm B)}_\sigma
  = \tfrac{1}{4}\Tr[\sigma\,(\SLDof{\sigma})^2]\,,
  \label{eq:Bures_metric_SLD}
\end{equation}
where the right-hand side is exactly one quarter of the standard QFI, i.e.
\begin{equation}
    \FQof{\sigma} =  \Tr[\sigma\,(\SLDof{\sigma})^2]\,.
    \label{eq:QFI_Def}
\end{equation}
Hence, the QFI $\FQof{\sigma}$
appears naturally as the local Bures metric (up to a multiplicative factor) of the statistical model,
representing the squared infinitesimal distance between neighbouring states
$\sigma(\beta)$ and $\sigma(\beta+d\beta)$ and thus quantifying the distinguishability of nearby states
and the sensitivity of the model to changes in the parameter~$\beta$. 

With this in mind and in order to assess the effect of unitary driving on local distinguishability, our goal is to compute the QFI of $\rho(t,\beta)$ according to \cref{eq:QFI_Def}.

We recall the main assumptions concerning the dynamics under consideration:
\begin{enumerate}
    \item At time $t=0$, the state of the thermometer is thermal (at inverse temperature $\beta$) with respect to its bare Hamiltonian, as a result of thermalization with a sample (whose temperature we want to estimate) that has then been subsequently disconnected from the probe. Thus, the initial manifold is that of Gibbs states $\{\piZero\}$, equipped with QFI $\FQof{\pi_0} = \Var_{\pi_0}( H_0)$.
    \item At times $t>0$, the probe Hamiltonian is coupled to an external perturbation $V$, yielding $H(t,\beta) = H_0 +\lamt V$, with $\lamt$ the driving profile, assumed to be temperature-dependent in the most general setting. The ensuing dynamics is $\rho(t,\beta)=\Ut\piZero\Utd$, with $\Ut = \mathcal{T}\exp\!\left[-\frac{i}{\hbar}\!\int_0^t\! H(s,\beta)\,ds\right]$.
    \teal{We remark that the particular additive structure for $H(t,\beta)$ is chosen in our work as the simplest ''response-theoretic" setting that allows one to isolate, within a fully unitary framework, the effect of a deterministic drive applied after thermalization. At the same time, this class of Hamiltonians captures a broad set of physically relevant driving scenarios: the decomposition into a drift term $H_0$ plus externally programmable control contributions is common in controlled quantum metrology and open quantum systems theory, where external fields couple through fixed system operators with time-dependent amplitudes.}
    
\end{enumerate}

First, notice that, exploiting the unitary covariance of $\JJ{\rho(t)}$ and its inverse, one can compute the SLD at time $t$, appearing in the definition of the QFI,
from its expression at the reference state $\pi_0$.

Differentiating $\rho(t,\beta)$ with respect to $\beta$ yields
\begin{equation}
  \partial_\beta\rho(t,\beta)
  = \Ut\Bigl(\partial_\beta\pi_0 + [A(t,\beta),\,\pi_0]\Bigr)\Utd,
  \qquad
  A(t,\beta):=\Utd\,\partial_\beta \Ut,
  \label{eq:drho_split}
\end{equation}
which separates the equilibrium derivative $\partial_\beta\pi_0$
from the dynamical contribution generated by $A(t,\beta)$. 
Then, from \cref{eq:Bures_covariance},
\begin{equation}
  \SLDof{\rho(t)}
  = \Ut\!\left[
      \JJi{\pi_0}\!(\partial_\beta\pi_0)
      + \JJi{\pi_0}\!([A(t,\beta),\pi_0])
    \right]\!\Utd,
  \label{eq:SLD_prelim}
\end{equation}
or equivalently
\begin{equation}
  \SLDof{\rho(t)}
  = \Ut \bigl( \SLDof{\pi_0} + \delta L^\beta_t\bigr) \Utd\;,
  \label{eq:SLD_split}
\end{equation}
having distinguished the equilibrium baseline $\SLDof{\pi_0} = - (H_0 -\Tr[\piZero H_0])$ from the dynamical correction
\begin{equation}
\delta L^\beta_t:=\JJi{\pi_0}\!\big([A(t,\beta),\pi_0]\big)\;.
\label{eq:deltaL_general_def}
\end{equation}

We thus need to compute the explicit form of $A(t,\beta)$, which requires differentiating $\Ut$.
The Dyson expansion of the unitary propagator reads
\begin{equation}
  \Ut = U(t,0;\beta)= \Id
  + \sum_{n=1}^{\infty}\!\left(-\frac{i}{\hbar}\right)^{\!n}
    \!\!\!\int_{0<t_1<\cdots<t_n<t}\!\!\!\!dt_1\cdots dt_n\;
    H(t_n,\beta)\cdots H(t_1,\beta),
  \label{eq:Dyson}
\end{equation}
where $t_n$ is the latest and $t_1$ the earliest time variable.
Differentiating each term of \cref{eq:Dyson} with respect to $\beta$ gives
\begin{align}
  \partial_\beta U(t,0;\beta)
  &= \sum_{n=1}^{\infty}\!\left(-\frac{i}{\hbar}\right)^{\!n}
     \!\!\!\int_{0<t_1<\cdots<t_n<t}\!\!\!\!dt_1\cdots dt_n
     \sum_{k=1}^{n}
     \Biggl[
     \Bigl(\prod_{r=k+1}^{n} H(t_r,\beta)\Bigr)
     (\partial_\beta H(t_k,\beta))
     \Bigl(\prod_{r=1}^{k-1} H(t_r,\beta)\Bigr)
     \Biggr],
  \label{eq:dU_sum}
\end{align}
with the convention
\begin{equation}
  \prod_{r=a}^{b} X_r := \Id \qquad \text{whenever } a>b,
  \label{eq:emptyproduct}
\end{equation}
so that the terms $k=1$ and $k=n$ are automatically included.

For a fixed $k$, we set $s:=t_k$ and split the integration variables
into the subsets $\{t_j>s\}$ and $\{t_j<s\}$.
Recognizing the corresponding partial propagators,
\begin{align}
  U(t,s;\beta)
  &= \Id + \sum_{p=1}^{\infty}\!\left(-\frac{i}{\hbar}\right)^{\!p}
     \!\!\!\int_{s<t_{k+1}<\cdots<t_n<t}\!\!\!\!dt_{k+1}\cdots dt_n\;
     \prod_{r=k+1}^{n} H(t_r,\beta), \nonumber\\[2pt]
  U(s,0;\beta)
  &= \Id + \sum_{q=1}^{\infty}\!\left(-\frac{i}{\hbar}\right)^{\!q}
     \!\!\!\int_{0<t_1<\cdots<t_{k-1}<s}\!\!\!\!dt_1\cdots dt_{k-1}\;
     \prod_{r=1}^{k-1} H(t_r,\beta),
  \label{eq:Usegments}
\end{align}
and summing over $k$ we recover
\begin{equation}
  \partial_\beta U(t,0;\beta)
  = -\frac{i}{\hbar}\int_0^t
     U(t,0;\beta)\,U^\dagger(s,0;\beta)\,\partial_\beta H(s,\beta)\,U(s,0;\beta)\,ds.
  \label{eq:dU_integral}
\end{equation}

Right-multiplying \cref{eq:dU_integral} by $U^\dagger(t,0,\beta)$ then yields
\begin{equation}
  A(t,\beta)= -\frac{i}{\hbar}\int_0^t U^\dagger_{s,\beta}\,(\partial_\beta H(s,\beta))\,U_{s,\beta}\,ds.
  \label{eq:X_identity}
\end{equation}
\teal{~\cref{eq:X_identity} is completely general and does not require any assumption on the specific form of the $\beta$-dependence of $H(t,\beta)$. Combining~\cref{eq:SLD_split,eq:deltaL_general_def} with the QFI definition (i.e. ~\cref{eq:QFI_Def}) yields
\begin{equation}
\FQof{\rho(t)}=\Tr\!\left[\piZero\left(\SLDof{\pi_0}+\delta L_t^\beta\right)^2\right]
=\underbrace{\Tr\left[\piZero\,\left(\SLDof{\pi_0}\right)^{2}\right]}_{\FQof{\pi_0}}
     \;+\;2\Tr\!\left[\piZero \SLDof{\pi_0}\delta L_t^\beta\right]\;+\;\underbrace{\Tr\!\left[\piZero\left(\delta L_t^\beta\right)^2\right]}_{\It}\;.
\label{eq:general_decomposition_statement}
\end{equation}
The mixed term vanishes. Indeed
\begin{align}
    \Tr\!\left[\piZero \SLDof{\pi_0}\delta L_t^\beta\right] &=  \Tr\!\left[\piZero \SLDof{\pi_0}\JJi{\pi_0}\!\big([A(t,\beta),\pi_0]\big)\right] =\\ \nonumber &=\Tr\!\left[\JJi{\pi_0}\!\big(\piZero \SLDof{\pi_0}\big)[A(t,\beta),\pi_0]\right] = \Tr\!\left[\SLDof{\pi_0}[A(t,\beta),\pi_0]\right]=0
\end{align}
where we use the cyclicity of the trace and the fact that $\JJi{\pi_0}(X) = \pi_0^{-1} X$ whenever $[X,\pi_0]=0$. The last equality follows directly from $\SLDof{\pi_0}$ commuting with $\pi_0$ whereas $[A(t,\beta),\pi_0]$ is off-diagonal in the equilibrium eigenbasis, so that their product has vanishing trace.
Conversely the quadratic term $\Tr\!\left[\piZero\left(\delta L_t^\beta\right)^2\right]$, is identified as the increment $\It$, and is manifestly non-negative, being the expectation value of a positive operator. 
This leads directly to Eq.~(1) of the main text, in full generality, for arbitrary $\beta$-dependent unitary dynamics. Before moving on, it is worth stressing that this decomposition of the QFI (baseline + non-negative increment) fundamentally relies on the initial state being full rank (ensuring the invertibility of $\JJ{\pi_0}$) and satisfying the commuting-derivative condition $[\pi_0(\beta),\partial_\beta \pi_0(\beta)]=0$ (which enables the cancellation of the mixed term in~\cref{eq:general_decomposition_statement}). These conditions are naturally satisfied, though not uniquely, by the manifold of canonical Gibbs states, further justifying this choice as a standard starting point for our analysis.}

We now specialize to the additive driving 
\begin{equation}
H(t,\beta)=H_0+\lamt V,
\end{equation}
for which $\partial_\beta H(t,\beta) = \partial_\beta \lamt\, V$. Substituting into \cref{eq:X_identity} gives
\begin{equation}
  A(t,\beta)
  = -\frac{i}{\hbar}\int_0^t (\partial_\beta\lambda(s,\beta))\,V_H(s)\,ds,
  \label{eq:A_additive}
\end{equation}
where $V_H(s):=U^\dagger(s,\beta)\,V\,U(s,\beta)$ is the Heisenberg-picture time-evolved perturbation.

Substituting \cref{eq:A_additive} into \cref{eq:deltaL_general_def}
and using the linearity of $\JJi{\pi_0}$ gives
\begin{equation}
  \delta L_t^\beta 
  = \int_0^t (\partial_\beta\lambda(s,\beta))\,
    J_V(s)\,ds,
  \label{eq:Ltv_final}
\end{equation}
having defined the information current operator 
\begin{equation}
  J_V(s):= -\frac{i}{\hbar}\,
  \JJi{\pi_0}\!\bigl([V_H(s),\pi_0]\bigr).
\end{equation}
Notice that, if $[V_H(s),\pi_0]=0$ at all times, then $J_V(s)=0$ and hence $\delta L_t^\beta=0$:
no dynamical correction to the SLD arises, and the QFI remains frozen at its equilibrium value.
Consistently, in the eigenbasis of $\pi_0$, the commutator $[V_H(s),\pi_0]$ has only off-diagonal
matrix elements, and the inverse superoperator $\JJi{\pi_0}$ preserves this structure.
Therefore $J_V(s)$ acts entirely within the operator subspace orthogonal to
$\piZero$, i.e. it is associated with the generation of coherences in the energy
eigenbasis of the thermal state.
Using the newly introduced information current, we express the QFI increment as 
\begin{align}
  \It=
     \int_0^t\!\!\int_0^t
        (\partial_\beta\lambda(s,\beta))(\partial_\beta\lambda(u,\beta))\,
        K(s,u,\beta)\,ds\,du\,\,,
  \label{eq:FQ_final}
\end{align}
with kernel
\begin{equation}
K(s,u,\beta)\;:=\;\Tr\!\bigl[\piZero\,J_V(s)\,J_V(u)\bigr]\,.
  \label{eq:K_def}
\end{equation}
Recovering Eq.~(13) of the main text.
As a final note, since $(\partial_\beta\lambda(s,\beta))(\partial_\beta\lambda(u,\beta))$ is symmetric in $(s,u)$,
only the Hermitian (symmetric) part of the kernel contributes, namely
\begin{equation}
  \int_0^t\!\!\int_0^t
  (\partial_\beta\lambda(s,\beta))(\partial_\beta\lambda(u,\beta))\,
  K(s,u,\beta)\,ds\,du
  \;=\;
  \int_0^t\!\!\int_0^t
  (\partial_\beta\lambda(s,\beta))(\partial_\beta\lambda(u,\beta))\,
  K_S(s,u,\beta)\,ds\,du,
\end{equation}
where 
\begin{equation}
  K_S(s,u,\beta):=\;\frac{1}{2}\,\Tr\!\bigl[\piZero\,\{J_V(s),J_V(u)\}\bigr]
  \;=\;\frac{1}{2}\bigl(K(s,u,\beta)+K(u,s,\beta)\bigr),
  \label{eq:Ks_def}
\end{equation}
while the antisymmetric part
\begin{equation}
    K_A(s,u,\beta) := \frac{1}{2} \bigl(K(s,u,\beta) - K(u,s,\beta)\bigr)
\end{equation}
does not contribute, i.e.
\begin{equation}
    \int_0^t\!\!\int_0^t
  (\partial_\beta\lambda(s,\beta))(\partial_\beta\lambda(u,\beta))\,K_A(s,u,\beta)\,ds\,du =0.
\end{equation}
\section*{S2: On the physical origin and operational interpretation of the temperature-dependent driving}
\label{app:S2}
\teal{In this section, we clarify the physical origin and operational interpretation of the temperature dependence entering the drive profile $\lamt$ in our analysis. 
First, we point out that allowing the driving Hamiltonian to depend on $\beta$ corresponds to a wide class of driving mechanisms, encompassing standard unitary-control scenarios, in which the driving field is fully externally programmed and independent of temperature, but also capturing all those scenarios in which the perturation acting on the probe is mediated, renormalized or biased by physical mechanisms whose response depends on temperature, so that the effective drive inherits a $\beta$-dependence from the underlying physics.
Throughout, it is convenient to keep in mind the factorized parametrization used in the main text,
\begin{equation}
\lambda(t,\beta)=\lambda_0\,G(\beta)\,f(t),
\label{eq:lambda_factorized_S2}
\end{equation}
where $f(t)$ is an externally controlled temporal profile, while $G(\beta)$ plays the role of the response function encoding how the actuator (i.e. the physical system implementing the drive) depends on temperature.
We now discuss representative scenarios in which such a response may arise.}

\subsection{Physical mechanisms leading to temperature dependence}

\teal{A first microscopic route to motivate the emergence of $\lamt$ is provided by the so-called work-reservoir (WR) approximation. After thermalization with the sample, one can indeed imagine that the unitary control to which the probe is subjected is generated by an auxiliary macroscopic system acting only as a classical work source. 
The key assumption in this scenario is that this auxiliary system is itself in equilibrium at inverse temperature $\beta$ and is sufficiently large (i.e., in the thermodynamic limit) that fluctuations of its relevant observables are, for all intents and purposes, negligible during the driving stage. In this regime, the reservoir exchanges energy coherently with the probe, whose reduced dynamics is thus well approximated by a unitary evolution.
To build intuition on this mechanism, consider, as an illustrative example, a longitudinal Ising chain of $N$ spins
\begin{equation}
H_{\mathrm{WR}}
=
-h\sum_{j=1}^N \sigma_j^z
-
J\sum_{j=1}^{N-1}\sigma_j^z\sigma_{j+1}^z ,
\end{equation}
and define the intensive magnetization
\begin{equation}
M_N=\frac{1}{N}\sum_{j=1}^N\sigma_j^z .
\end{equation}
Suppose now that the probe is coupled to this observable through
\begin{equation}
H_{\mathrm{int}}(t)
=
\lambda_0 f(t)\,V\otimes M_N .
\end{equation}
The total Hamiltonian during the driving stage is therefore
\begin{equation}
H_{\mathrm{tot}}(t)=H_0+H_{\mathrm{WR}}+H_{\mathrm{int}}(t).
\end{equation}
Assuming an initially factorized state
\begin{equation}
\rho_{\mathrm{tot}}(0,\beta)
=
\pi_0(\beta)\otimes\tau_{\mathrm{WR}}(\beta) \qquad \text{with \,\, }\tau_{\mathrm{WR}}(\beta)
=
\frac{e^{-\beta H_{\mathrm{WR}}}}{\mathrm{Tr}[e^{-\beta H_{\mathrm{WR}}}]}\;,
\end{equation}
the probe-only dynamics can be written as a convex mixture of unitary evolutions generated by Hamiltonians $H_0+\lambda_0 m f(t)V$, where $m$ are the eigenvalues of $M_N$. In the thermodynamic limit and away from criticality (trivial for a 1D Ising chain at finite temperature), however, the magnetization density becomes self-averaging~\cite{book:Balian}, with fluctuations vanishing as $N\to\infty$, and its distribution sharply peaked around the equilibrium expectation value
\begin{equation}
\langle M_N\rangle_\beta
=
\mathrm{Tr}[M_N\tau_{\mathrm{WR}}(\beta)]\,.
\end{equation}
Consequently, the above-mentioned convex mixture collapses to an effectively deterministic unitary propagator generated by
\begin{equation}
H_{\mathrm{eff}}(t,\beta)
=
H_0+\lambda_0\langle M_N\rangle_\beta f(t)V .
\end{equation}
Comparing with \cref{eq:lambda_factorized_S2}, it is thus immediate to identify
\begin{equation}
G(\beta)=\langle M_N\rangle_\beta \,,
\end{equation}
showing how the temperature dependence of the perturbations follows directly from the equilibrium properties of the WR.}\\

\teal{Beyond this mechanism, an alternative microscopic route leading to an effective temperature-dependent perturbation of the probe dynamics arises naturally from general weak coupling of the probe to auxiliary degrees of freedom.
Indeed, in standard treatments of Markovian open quantum dynamics, the reduced probe evolution is described by a Gorini–Kossakowski–Sudarshan–Lindblad master equation
\begin{equation}
\dot{\rho}(t)
=
-\frac{i}{\hbar}[H_0+H_{\mathrm{LS}}(\beta),\rho(t)]
+
\mathcal{D}_\beta(\rho(t)) ,
\end{equation}
where $H_{\mathrm{LS}}(\beta)$ is the Lamb-shift Hamiltonian, contributing as a coherent correction to the probe's dynamics, while $\mathcal{D}_\beta$ encodes dissipative effects. Both terms originate from environmental correlation functions and therefore may inherit an explicit temperature dependence when the auxiliary degrees of freedom are themselves in a thermal state at the true inverse temperature $\beta^\ast$.
Although the resulting dynamics is not strictly unitary, in the short-time and weak-dissipation regime, the Lamb-shift term represents the leading-order correction to the isolated dynamics, while dissipative effects become appreciable only at higher orders in time. For short interrogation times ( i.e. $t \ll \gamma^{-1}$, with $\gamma$ denoting the characteristic dissipative rates appearing in $\mathcal{D}$), the reduced dynamics remains perturbatively close to unitary evolution and can therefore be consistently approximated by an effective Hamiltonian description with an explicit temperature-dependent term. Indeed, the relevance of Hamiltonian (Lamb-shift–type) contributions to non-equilibrium thermometric gain has already been observed in the literature (e.g. in~\cite{art:Sekatski_Optimal}).
This further suggests that our results can also be interpreted as a model-independent baseline extending beyond the fully unitary assumption: although the non-negativity of the increment might not (and in general will not) survive arbitrary noise, the unitary-induced increment identifies a natural upper bound for 
the maximum attainable sensitivity when the implemented control remains perturbatively close to unitary, before dissipative channels may dampen its effect.}\\

\teal{As a final note, besides these microscopic models, we remark that one may simply regard $G(\beta)$ as an effective response function associated with temperature-dependent material properties entering the control apparatus itself. Typical examples include the thermo-optic effect, whereby the refractive index of a material depends on temperature, leading, for example, to shifts of the supported mode frequencies in a resonant optical cavity, or similar temperature-dependent drifts in the behavior of electronic control components. In all such cases the experimentally implemented field remains externally shaped in time, but its effective amplitude or detuning acquires a residual dependence on temperature, justifying the formalism developed in the main text.}

\subsection{Operationality of the control}

\teal{With the physical origin of the temperature dependence clarified, we now discuss how such a control is handled within the estimation protocol.}

\teal{Our analysis is naturally formulated within the framework of local quantum estimation theory, where one universally assumes that a preliminary coarse estimate $\beta_0$ of the parameter is available a priori and the protocol is designed and optimized to operate in a neighborhood of that value. Within this framework, the experimenter is further assumed to have direct control over the temporal waveform $f(t)$, which can be engineered to achieve a desired dynamical condition (for example resonance with the probe in the single spin-$1/2$ scenario), while also being able to shift and bias the operating region of  $G(\beta)$ by selecting the working point $\beta_0$. Such a working point may be obtained, for example, by devoting part of the available measurement budget (e.g. a number of experimental repetitions) to a preliminary equilibrium estimation protocol. In this sense, the parameters entering $G(\beta)$ should be regarded as design parameters fixed before the refined estimation stage.}

\teal{A natural question is therefore how precise this prior estimate must be for the driving protocol to yield a significant increment in the QFI. In other words, how sensitive is the dynamical enhancement $I_\beta^t$ to a mismatch between the assumed working point $\beta_0$ and the true temperature $\beta$? 
Now, for drivings such as \cref{eq:lambda_factorized_S2}, the dynamical increment reads
\begin{equation}
I_\beta^t
=
\lambda_0^2 [G'(\beta)]^2
\int_0^t\!\!\int_0^t
f(s)f(u)\,
K_S(s,u,\beta)\,ds\,du .
\label{eq:I_factorized_SM}
\end{equation}
Hence, up to the probe-dependent kernel, the enhancement is governed by the response factor $[G'(\beta)]^2$, which quantifies how efficiently small temperature variations are converted into changes of the effective driving. The protocol is therefore expected to be most effective in regions where the response function has an appreciable slope, namely where small temperature changes translate into sizable modifications of the effective drive: the enhancement is thus most significant whenever the  prior-informed working region of parameter space (whose size is dictated by the precision of $\beta_0$) overlaps with the interval of maximal variation for $G(\beta)$.
To illustrate this mechanism, consider the Gaussian shape discussed in the main text
\begin{equation}
G(\beta)
=
\exp\!\left[-\frac{(\beta-\beta_0)^2}{2s_\beta^2}\right].
\label{eq:GaussianG_S2}
\end{equation}}

\teal{We stress once again that the exact form of $G(\beta)$ depends on the physics of the actuator providing the drive, and that our Gaussian choice has no deep microscopic motivation. Rather, it acts as a simple and transparent illustrative proxy for a generic smooth response localized around a chosen working region, without committing to any specific mechanism: the qualitative considerations derived from it remain valid even for more specific $G(\beta)$ profiles. For the Gaussian, one finds
\begin{equation}
[G'(\beta)]^2
=
\frac{(\Delta\beta)^2}{s_\beta^4}
\exp\!\left[-\frac{(\Delta\beta)^2}{s_\beta^2}\right],
\end{equation}
where $\Delta\beta=\beta-\beta_0$.
This quantity vanishes at $\Delta\beta=0$ and reaches its maximum at $|\Delta\beta|=s_\beta$. Thus $\beta_0$ identifies the center of the response window, while the optimal operating region lies slightly off center, i.e. near $\beta_0\pm s_\beta$ where the slope $|G'|$ is largest (rather than exactly at the symmetric center).}

\begin{figure}[t]
    \centering
    \includegraphics[width=0.8\linewidth]{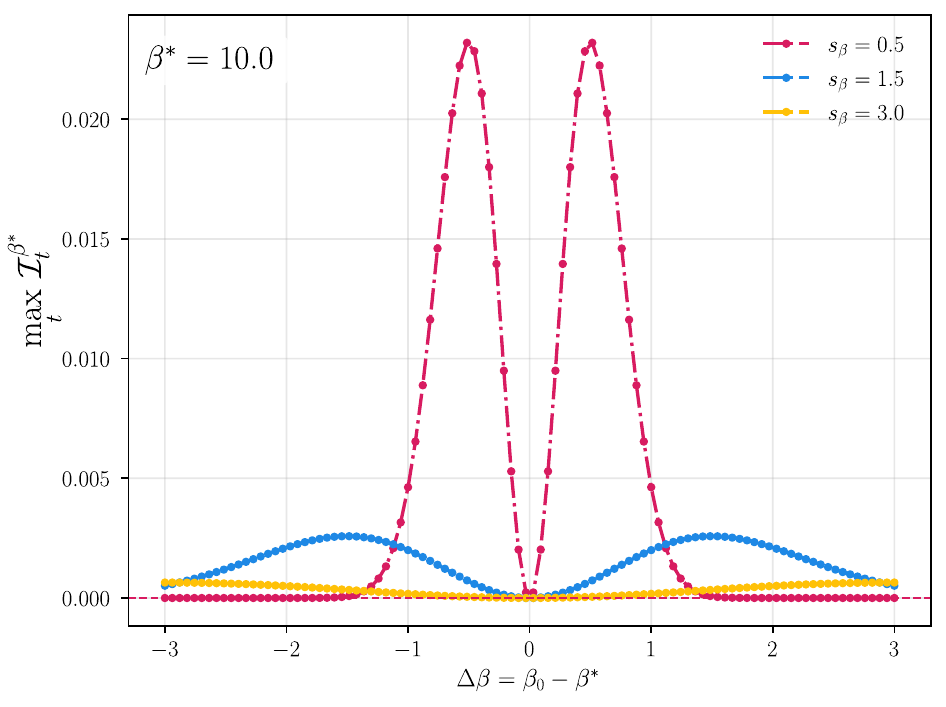}
    \caption{
    \teal{Driving-induced increment $I_\beta^t$ as a function of the working-point mismatch $\Delta\beta=\beta-\beta_0$ for three representative Gaussian widths $s_\beta$, with fixed $\beta^\ast=10.0$. The increment vanishes at $\Delta\beta=0$ due to the symmetry of the Gaussian envelope  and is maximal at $|\Delta\beta|\approx s_\beta$, in agreement with the analytic factor $(\Delta\beta)^2 e^{-(\Delta\beta)^2/s_\beta^2}$. Narrower envelopes yield larger peak enhancement but reduced robustness to detuning, requiring a more precise rough estimate to correctly position the working point and therefore a larger allocation of resources to the preliminary estimation stage. Conversely, broader envelopes improve tolerance to calibration mismatch at the expense of maximal gain.}
    }
    \label{fig:delta_beta_SM}
\end{figure}

\teal{This also highlights a clear robustness trade-off: as shown in~\cref{fig:delta_beta_SM}, narrower envelopes yield larger peak enhancement but only over a smaller temperature interval, making the protocol more sensitive to working-point mismatch. Conversely, broader envelopes reduce the maximal gain but improve tolerance to detuning from the nominal working point.
Operationally, this implies that if the regions where $G(\beta)$ varies most rapidly are narrow, positioning the working point within them requires a more precise preliminary estimate of the temperature. Achieving the largest dynamical enhancement may therefore require allocating a larger fraction of the available measurement budget to the rough estimation stage used to determine the working point. Conversely, broader response profiles relax this calibration requirement, at the cost of a smaller maximal enhancement.}

\FloatBarrier

\section*{S3: Single-spin probe}
\label{app:S3}
In this section, we discuss the results concerning the representative model of the driven spin probe. Our main goal is to instantiate Eq.~(13) of the main text for a single spin-$1/2$ under coherent driving, showing explicitly the behaviour of the QFI increment at short and long times. Throughout we set, for clarity, $\hbar=k_{\mathrm B}=1$.

We begin by computing the model-specific form of the increment $\It$ from \cref{eq:FQ_final}. 
At $t=0$ the single spin is in a Gibbs state with respect to $H_0=\frac{\Omega}{2}\sigma_z$ at inverse temperature $\beta$,
\begin{equation}
  \piZero=\frac{e^{-\beta H_0}}{Z_0}
  =\tfrac{1}{2}\bigl(\Id - m\,\sigma_z\bigr),
  \qquad
  m=\tanh\!\left(\frac{\beta\Omega}{2}\right),
  \label{eq:pi0_def}
\end{equation}
with $Z_0=\Tr[e^{-\beta H_0}]$ and $\Tr[\pi_0\sigma_z]=-m$.

For $t>0$ we apply a transverse control $V=\sigma_x$ modulated by a cosinusoidal driving profile, $\lamt = \lambda_0 G(\beta) \cos(\omega_{\mathrm d} t +\phi)$, so that
\begin{equation}
  H(t,\beta)=H_0+\lambda(t,\beta)\,V
  =\frac{1}{2}\,\mathbf B(t,\beta)\!\cdot\!\boldsymbol\sigma,
  \qquad
  \mathbf B(t,\beta)=\bigl(2\lambda_0 G(\beta)\cos(\omega_{\mathrm d}t+\phi),\,0,\,\Omega\bigr),
  \label{eq:H_Bloch}
\end{equation}
where $\boldsymbol\sigma=(\sigma_x,\sigma_y,\sigma_z)$, $\lambda_0$ is the drive amplitude, $G(\beta)$ a Gaussian temperature envelope, and $\omega_{\mathrm d}$ and $\phi$ the drive frequency and phase, respectively. We set, without loss of generality, $\phi=0$. 
Since every $\mathrm{SU}(2)$ unitary acts as a proper rotation on spin operators, the Heisenberg-picture time-evolved perturbation can always be expressed as 
\begin{equation}
  V_H(t)=\Utd\,V\,\Ut
  =\boldsymbol\sigma\!\cdot\!\mathbf a(t),
  \qquad
  \mathbf a(0)=(1,0,0),
  \label{eq:VH_def}
\end{equation}
with the Bloch vector $\mathbf a(t)=(a_x(t),a_y(t),a_z(t))$ obeying the precession equation
\begin{equation}
  \dot{\mathbf a}(t)=\mathbf B(t,\beta)\times\mathbf a(t).
  \label{eq:bloch_precession}
\end{equation}

As discussed in the previous section, the information current operator is defined as 
\begin{equation}
  J_V(t)=-i\,\JJi{\pi_0}\!\bigl([V_H(t),\pi_0]\bigr).
\end{equation}
Using \cref{eq:pi0_def} and $[\sigma_i,\sigma_j]=2i\,\epsilon_{ijk}\sigma_k$,
\begin{equation}
  [\boldsymbol\sigma\!\cdot\!\mathbf a(t),\pi_0]
  = -\frac{m}{2}\,[\,\boldsymbol\sigma\!\cdot\!\mathbf a(t),\sigma_z\,]
  = i\,m\bigl(a_x(t)\sigma_y - a_y(t)\sigma_x\bigr).
  \label{eq:comm_pi0}
\end{equation}
Consider now a generic operator $X$ that is fully off-diagonal in the eigenbasis of $\pi_0$, and can thus be written as a linear combination of $\sigma_x$ and $\sigma_y$.  
Using the integral representation of $\JJi{\pi_0}$ and \cref{eq:pi0_def}, we write
\begin{equation}
  \JJi{\pi_0}(X)
  =2\!\int_0^{\infty} e^{-s\pi_0}X e^{-s\pi_0}\,ds 
  =2\!\int_0^{\infty}e^{-s/2}\,e^{(sm/2)\sigma_z} X \,e^{-s/2}\,e^{(sm/2)\sigma_z}\,ds .
  \label{eq:Jinv_integral_def}
\end{equation}
Since $X$ is a linear combination of $\sigma_x$ and $\sigma_y$, it anticommutes with $\sigma_z$, and one verifies that
\begin{equation}
  e^{\alpha\sigma_z}Xe^{\alpha\sigma_z}=X
  \qquad \forall\,\alpha\in\mathbb R,
\end{equation}
so that
\begin{equation}
  e^{-s/2}\,e^{(sm/2)\sigma_z} X \,e^{-s/2}\,e^{(sm/2)\sigma_z}
  = e^{-s}\,X.
\end{equation}
Therefore
\begin{equation}
  \JJi{\pi_0}(X)
  =2\!\int_0^{\infty}\!e^{-s}\,ds\,X
  =2X,
  \label{eq:Jinv_offdiag_result}
\end{equation}
and the overall information current takes the form
\begin{equation}
  J_V(t)
  =2m\bigl(a_x(t)\sigma_y-a_y(t)\sigma_x\bigr).
\end{equation}

To compute the integration kernel $K(s,u,\beta)$ and its symmetrized form, we exploit $\sigma_i\sigma_j=\delta_{ij}\Id+i\epsilon_{ijk}\sigma_k$ to derive 
\begin{equation}
J_V(s)J_V(u)
=(2m)^2\!\Bigl[
\bigl(a_x(s)a_x(u)+a_y(s)a_y(u)\bigr)\Id
+i\bigl(a_x(s)a_y(u)-a_y(s)a_x(u)\bigr)\sigma_z
\Bigr].
\end{equation}
In turn, given $\Tr[\pi_0\,\Id]=1$ and the definition of $m$, 
\begin{equation}
    K(s,u,\beta)
=\Tr[\pi_0\,J_V(s)J_V(u)]
=4m^2\Bigl[
a_x(s)a_x(u)+a_y(s)a_y(u)
-i\,m\,\bigl(a_x(s)a_y(u)-a_y(s)a_x(u)\bigr)
\Bigr].
\label{eq:K_full}
\end{equation} 
Finally, the symmetrized kernel, corresponding to the Hermitian part of $K(s,u,\beta)$, reads
\begin{equation}
K_S(s,u,\beta)=\tfrac{1}{2}\Tr\!\bigl[\pi_0\{J_V(s),J_V(u)\}\bigr]
=4m^2\bigl[a_x(s)a_x(u)+a_y(s)a_y(u)\bigr].
\label{eq:Ks}
\end{equation}
We thus recover the exact, although in general involved, form of the increment at time $t$:
\begin{equation}
\label{eq:I_qubit_exact}
\It
=
\int_0^t\!\int_0^t\;\,
\partial_\beta\lambda(s,\beta)\,\partial_\beta\lambda(u,\beta)\,
4m^2\!\left[a_x(s)a_x(u)+a_y(s)a_y(u)\right]\,
\,ds\,du.
\end{equation}

\subsection{Short-time scaling}

We now specialize to the short-time behaviour of \cref{eq:I_qubit_exact}. To do so, we perform the standard reparametrization $s=t z$ and $u=t v$,
and expand the integrand in powers of the evolution time $t$.  

For our driving protocol $\partial_\beta\lambda(t,\beta)
=\lambda_0 G'(\beta)\cos(\omega_{\mathrm d} t)
=\lambda_0 G'(\beta)+O(t^2)$, where $G'(\beta) = \partial_\beta G(\beta)$. From the precession equation 
$\dot{\mathbf a}=\mathbf B\times\mathbf a$
with $\mathbf B(0)=(2\lambda_0 G(\beta),0,\Omega)$ and $\mathbf a(0)=(1,0,0)$,
we obtain, to leading order in $t$,
\begin{equation}
    a_x(t)=1+O(t^2),\qquad 
    a_y(t)=\Omega t+O(t^2)\;.
\end{equation}  
Hence, the symmetrized kernel admits the expansion
\begin{equation}
    K_S(s,u,\beta)=4m^2[a_x(ts)a_x(tu)+a_y(ts)a_y(tu)]
=4m^2[1+\Omega^2t^2su+O(t^3)].
\end{equation}
Substituting into 
\begin{equation}
    \It
= t^2
\int_0^1\int_0^1\!\;\,
\partial_\beta\lambda(ts,\beta)\,\partial_\beta\lambda(tu,\beta)\, K_S(ts,tu,\beta)\,
\,ds\,du
\end{equation}
gives
\begin{equation}
\mathcal I_t^\beta
=4m^2[\lambda_0 G'(\beta)]^2 t^2+O(t^4),
\end{equation}
revealing the expected leading quadratic dependence on time, as is typical for unitary dynamics with analytic parameter dependence around $t=0$.

\subsection{Long-time scaling in the weak-field regime}

To analyze the long-time behaviour of $\It$, we work under the weak-field assumption $\lambda_0\!\ll\!\Omega$, which makes the precession equation analytically tractable.
In this scenario, one can expand the Heisenberg-picture operator $V_H(t)$ around the interaction picture as
\begin{equation}
    V_H (t) \approx V_I(t) - i\int_0^t \lambda(s,\beta) [V_I(s),V_I(t)]\,ds + O(\lambda_0^2),
\end{equation}
with $V_I(t) = e^{iH_0t}V e^{-iH_0t} = \cos(\Omega t )\sigma_x -\sin(\Omega t) \sigma_y$ the interaction-picture evolved perturbation. Since $[V_I(s),V_I(t)]\propto \sigma_z$, the first-order correction vanishes when computing the commutator with $\piZero$ appearing in the information current. Then, up to first order in $\lambda_0$, we obtain 
\begin{equation}
    J_V(t)=2m\!\left(\cos(\Omega t)\,\sigma_y-\sin(\Omega t)\,\sigma_x\right)\;.
\label{eq:J_weakfield}
\end{equation}
In turn, the symmetrized kernel simplifies to 
\begin{equation}
K_S(s,u,\beta)=4\,m^2\,\cos\!\big(\Omega(s-u)\big)\;,
\label{eq:Ks_WeakField}
\end{equation}
having exploited the trigonometric identity $\cos(\Omega(s-u))=\cos(\Omega s)\cos(\Omega u)+\sin(\Omega s)\sin(\Omega u)$.

Inserting \cref{eq:Ks_WeakField} into the definition of the increment and using
$\partial_\beta\lambda(t,\beta)=\lambda_0 G'(\beta)\cos(\omega_{\mathrm d} t)$, we obtain to leading order in $\lambda_0$
\begin{align}
\mathcal I_t^{\beta}
&=\int_0^t\int_0^t\;
\partial_\beta\lambda(s,\beta)\,\,\partial_\beta\lambda(u,\beta)\,K_S(s,u,\beta)\, ds\,du \nonumber\\
&\approx4m^2\big[\lambda_0 G'(\beta)\big]^2
\int_0^t\int_0^t
\cos(\omega_{\mathrm d} s)\cos(\omega_{\mathrm d} u)\,
\cos\!\big(\Omega(s-u)\big)\,ds\,du\;.
\end{align}
Using once again the above-mentioned trigonometric identity, the double integral separates:
\begin{align}
\mathcal I_t^{\beta}
&\approx4m^2\big[\lambda_0 G'(\beta)\big]^2
\left[
\left(\int_0^t\!\cos(\omega_{\mathrm d} s)\cos(\Omega s)\,ds\right)^{\!2}
+\left(\int_0^t\!\cos(\omega_{\mathrm d} s)\sin(\Omega s)\,ds\right)^{\!2}
\right]\;.
\end{align}
Direct evaluation of each elementary integral gives
\begin{equation}
\mathcal I_t^{\beta}
\approx 4m^2\big[\lambda_0 G'(\beta)\big]^2 A(t),
\end{equation}
where 
\begin{equation}
    A(t):= \frac{\bigl(\omega_{\mathrm d} \sin(\omega_{\mathrm d} t)\cos(\Omega t) - \Omega \cos(\omega_{\mathrm d}t)\sin(\Omega t)\bigr)^2+\bigl(\omega_{\mathrm d} \sin(\omega_{\mathrm d} t)\sin(\Omega t) + \Omega \cos(\omega_{\mathrm d}t)\cos(\Omega t)-\Omega\bigr)^2}{(\omega_{\mathrm d}^2-\Omega^2)^2}\,\,.
\end{equation}
In the resonance limit $\omega_{\mathrm d} \to \Omega$, one finds
\begin{equation}
A(t)\to
\frac{2\Omega^2 t^2+2 \Omega t \sin(2\Omega t)-\cos(2\Omega t) +1}{8\Omega^2}\,\,.
\end{equation}
In the long-time limit, the oscillatory terms average out, so that the leading term remains proportional to $t^2$. In particular, we obtain
\begin{equation}
    \It \approx m^2\big[\lambda_0 G'(\beta)\big]^2 t^2,
\end{equation}
which is of the form
\begin{equation}
    \It \approx \bigl(\lambda_0 G'(\beta)\bigr)^2 \alpha(\beta)\,t^2,
\end{equation}
with $\alpha(\beta)=m^2$, as reported in Eq.~(17) of the main text. This confirms that the quadratic scaling holds even at long times in the resonant, weak-field regime.
\section*{S4: On the relation between Fisher increment and coherence-generation}
\label{app:S4}

\teal{In this section we relate the information-current kernel $K(s,u,\beta)$ to the
non-commutativity between the Heisenberg-evolved perturbation $V_H(s)$ and the
equilibrium state $\piZero$. The goal is to make precise, in a quantitative but
general way, that the generation of coherences in the eigenbasis of $\piZero$
is a necessary condition for a non-vanishing kernel and thus for a
nonzero dynamical increment $\It$.
Throughout this section we use the Schatten-$p$ norms  ($\|X\|_p :=
(\Tr|X|^p)^{1/p}$ for $p\ge1$) of the commutator as a quantifier of operator non-commutativity, a standard choice in the study of commutator bounds and matrix inequalities
(e.g.,~\cite{art:Bottcher, art:Wenzel, art:Cheng,art:Kittaneh}).
We focus in particular on $\|X\|_2$ , i.e.  the Hilbert--Schmidt
norm, obeying the standard the Cauchy-Schwarz inequality 
\begin{equation}
|\Tr(A^\dagger B)| \le \|A\|_2\,\|B\|_2.
\label{eq:HS-CS}
\end{equation}
Recall the kernel defined in \cref{eq:K_def}
\begin{equation}
K(s,u,\beta) := \Tr\!\big[\piZero\, J_V(s)\, J_V(u)\big].
\label{eq:kernel-def-S3}
\end{equation}
For notational clarity, we define the auxiliary operators
\begin{equation}
B_s := \pi_0^{1/2} J_V(s),\qquad B_u := \pi_0^{1/2} J_V(u).
\label{eq:BsBu-def}
\end{equation}
By cyclicity of the trace,
\begin{equation}
K(s,u,\beta)=\Tr\!\big[(\pi_0^{1/2}J_V(s))^\dagger\,(\pi_0^{1/2}J_V(u))\big]
           =\Tr\!\big[B_s^\dagger B_u\big],
\label{eq:kernel-BsBu}
\end{equation}
where we also used that $J_V(t)$ is Hermitian. Applying \cref{eq:HS-CS} yields the
general estimate
\begin{equation}
|K(s,u,\beta)| \le \|B_s\|_2\,\|B_u\|_2,
\qquad
\|B_t\|_2^2=\Tr\!\big[J_V(t)\,\piZero\,J_V(t)\big].
\label{eq:kernel-CS-bound}
\end{equation}
Thus, controlling $|K(s,u,\beta)|$ reduces to controlling the weighted
Hilbert-Schmidt size of the current $J_V(t)$.}

\teal{We now connect $\|B_s\|_2$ to a Schatten norm of the commutator
$[V_H(s),\piZero]$. Introduce
\begin{equation}
C(s):=[V_H(s),\piZero].
\label{eq:C-def}
\end{equation}
In the eigenbasis of $\piZero = \sum_k p_k\,\ket{k}\!\bra{k}$ the commutator has matrix elements
\begin{equation}
C_{kl}(s) = (p_l-p_k)\,\bra{k}V_H(s)\ket{l},
\label{eq:C-elements}
\end{equation}
so in particular $C_{kk}(s)=0$: $C(s)$ is purely off-diagonal in the eigenbasis
of $\piZero$, i.e.\ it directly captures coherence generation with respect to
that basis.}

\teal{Next, recall the definition of the information current,
\begin{equation}
J_V(s)= -\frac{i}{\hbar}\,\JJi{\pi_0}\!\bigl(C(s)\bigr).
\label{eq:JV-def-S3}
\end{equation}
In the $\piZero$ eigenbasis, given the expression of the inverse Bures-Jordan superoperator, we have 
\begin{equation}
\big(\JJi{\pi_0}(X)\big)_{kl}=\frac{2}{p_k+p_l}\,X_{kl}\implies \big(J_V(s)\big)_{kl}
= -\frac{i}{\hbar}\,\frac{2}{p_k+p_l}\,C_{kl}(s).
\label{eq:JV-elements}
\end{equation}
Conversely, computing $\|B_s\|_2^2$ yields 
\begin{align}
\|B_s\|_2^2
&=\Tr\!\big[B_s^\dagger B_s\big]
 =\Tr\!\big[J_V(s)\,\piZero\,J_V(s)\big]
 =\sum_{k,l} p_k\,\big|\big(J_V(s)\big)_{kl}\big|^2 = \frac{4}{\hbar^2}\sum_{k,l} p_k\,\frac{|C_{kl}(s)|^2}{(p_k+p_l)^2} .
\label{eq:Bs-norm-expanded}
\end{align}
Bounding the population factor by the state-dependent constant
\begin{equation}
\frac{p_k}{(p_k+p_l)^2}\le
\max_{i,j}\frac{p_i}{(p_i+p_j)^2}
=:c(\pi_0),
\label{eq:population-filter-bound}
\end{equation}
we obtain
\begin{equation}
\|B_s\|_2^2
\le \frac{4}{\hbar^2}\,c(\pi_0)\sum_{k,l}|C_{kl}(s)|^2
= \frac{4}{\hbar^2}\,c(\pi_0)\,\|C(s)\|_2^2
= \frac{4}{\hbar^2}\,c(\pi_0)\,\|[V_H(s),\piZero]\|_2^2.
\label{eq:Bs-norm-comm-bound}
\end{equation}
Taking square roots and combining with \cref{eq:kernel-CS-bound} yields the final bound
\begin{equation}
|K(s,u,\beta)|
\le
\mathbb C(\pi_0)\,
\|[V_H(s),\piZero]\|_2\,
\|[V_H(u),\piZero]\|_2,
\label{eq:kernel-final-bound}
\end{equation}
with 
$\mathbb C(\pi_0):=\frac{4}{\hbar^2}\,c(\pi_0)$}

\teal{\cref{eq:kernel-final-bound} makes explicit that if
$[V_H(s),\piZero]=0$ at all times, then $K(s,u,\beta)=0$ for all $(s,u)$ and the
increment $\It$ vanishes. In this precise sense, coherence generation in the
eigenbasis of $\piZero$ is a necessary condition for nonequilibrium
enhancement.}

\teal{At the same time, \cref{eq:kernel-final-bound} is generally not tight: the
population filter $p_k/(p_k+p_l)^2$ in \cref{eq:Bs-norm-expanded} shows that only
selected matrix elements contribute appreciably, depending on the thermal
spectrum. Moreover, the increment $\It$ involves a two-time integral of the
kernel weighted by the protocol-dependent factors $\partial_\beta\lambda(s,\beta)$
and $\partial_\beta\lambda(u,\beta)$; hence temporal correlations can suppress
the net gain even when the instantaneous commutator norm is large. Accordingly, the bound should be read as a general compatibility constraint rather than as a tight quantitative predictor of $\It$: non-commutativity is necessary for a nonzero enhancement, but its magnitude alone does not determine the size of the increment.}

\section*{S5: Resource accounting - gain versus control cost}
\label{app:S5}

\teal{In this section we benchmark the nonequilibrium metrological gain $\It$ against a conservative figure of merit for the
implemented control effort, thus making explicit a gain-cost balance for our protocol. The notion of
``energetic cost'' is not unique in driven quantum systems, and can depend strongly on the physical
realization and on the adopted operational definition of work. Here we adopt a control-theoretic viewpoint
and quantify cost through a norm functional of the externally applied Hamiltonian term.
This choice is directly motivated by the shortcut-to-adiabaticity / counterdiabatic driving literature.
A standard approach emphasized in this context is to quantify resources through time-integrated norms of the
implemented auxiliary Hamiltonian, with the precise functional depending on the experimental platform
\cite{art:CampbellTradeoffSTA,art:ZhengCostCD}. This is also the structural choice made in many quantum optimal-control
formulations, where quadratic field penalties are introduced to suppress excessively strong controls and
to encode amplitude/fluence constraints in a setup-agnostic way \cite{art:WerschnikQOCT}.}

\teal{In the same spirit, we use as a cost proxy a norm functional of the externally generated control term. In our setting the latter is
\begin{equation}
H_{\rm ctrl}(t,\beta)=\lambda(t,\beta)\,V.
\end{equation}
We introduce the time-integrated squared Hilbert--Schmidt (Frobenius) norm,
\begin{equation}
  \mathcal{C}_2(t,\beta)
  :=\int_0^t ds\;\|H_{\rm ctrl}(s,\beta)\|_2^2,
  \label{eq:C2_def_S5}
\end{equation}
which is an explicitly state-independent measure of the applied field strength over the protocol. For
$H_{\rm ctrl}=\lambda V$ it reduces to
\begin{equation}
  \mathcal{C}_2(t,\beta)=\|V\|_2^{\,2}\int_0^t ds\;|\lambda(s,\beta)|^2.
  \label{eq:C2_lambda_S5}
\end{equation}
This should be contrasted with energetic diagnostics such as the average injected work,
\begin{equation}
  W_t=\int_0^t ds\;\Tr\!\big[\rho(s,\beta)\,\partial_s H(s,\beta)\big]
     =\int_0^t ds\;\dot\lambda(s,\beta)\,\Tr\!\big[\rho(s,\beta)\,V\big],
  \label{eq:work_def_S5}
\end{equation}
which is explicitly state-dependent through $\Tr[\rho(s,\beta)V]$ and therefore can be strongly protocol-
and model-dependent. Since our goal here is to provide a conservative, setup-agnostic benchmark that depends
only on the applied control term, we focus on \cref{eq:C2_def_S5}.
Given this cost metric, we quantify ``information per unit control'' through the ratio
\begin{equation}
  \mathcal{R}_2(t,\beta):=\frac{\It}{\mathcal{C}_2(t,\beta)}.
  \label{eq:R2_def_S5}
\end{equation}
For factorized drivings, $\lambda(t,\beta)=\lambda_0\,G(\beta)\,f(t)$, \cref{eq:C2_lambda_S5} gives
\begin{equation}
  \mathcal{C}_2(t,\beta)=\|V\|_2^{\,2}\,\lambda_0^2\,G(\beta)^2\int_0^t ds\;f(s)^2.
  \label{eq:C2_factorized_S5}
\end{equation}
On the gain side, the increment depends on $\partial_\beta\lambda(t,\beta)=\lambda_0\,G'(\beta)\,f(t)$
(see \cref{eq:I_factorized_SM}), hence $\It\propto \lambda_0^2 [G'(\beta)]^2$ at the level of explicit
prefactors. As a result, $\mathcal{R}_2$ is independent of the overall drive scale $\lambda_0$:
this figure of merit cannot be improved by simply ``ramping up'' the control amplitude, but is instead governed
by (i) the temperature responsivity through the dimensionless factor $[G'(\beta)/G(\beta)]^2=[\partial_\beta\ln G(\beta)]^2$,
and (ii) the temporal structure through the ratio of the corresponding time integrals.}

\begin{figure}[t]
  \centering
  \includegraphics[width=0.95\linewidth]{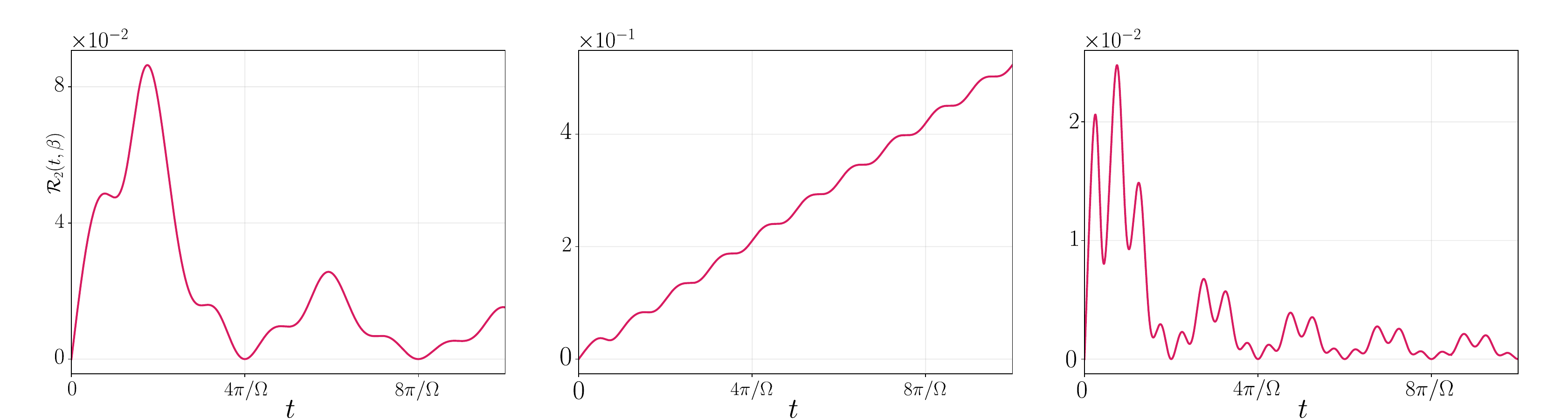}
  \caption{\teal{Gain-cost comparison for the driven single spin-$1/2$ model. The ratio 
$\mathcal{R}_2(t,\beta)=I_\beta^t/\mathcal{C}_2(t,\beta)$ is plotted as a function of time $t$ (expressed in units of the spin characteristic period $2\pi/\Omega$, with $\Omega = 1$)
for different driving frequencies: $\omega_d/\Omega=0.5$ 
(below resonance), $\omega_d/\Omega=1$ (on resonance), and 
$\omega_d/\Omega=2.0$ (above resonance). All driving profile parameters are chosen to be the same as in 
Fig.~(2) of the main text.}}
  \label{fig:cost_tradeoff}
\end{figure}

\teal{We now specialize to the driven single spin-$1/2$ probe, where $V=\sigma_x$ and $\|V\|_2^2=\Tr(\sigma_x^2)=2$, and
$\lambda(t,\beta)=\lambda_0\,G(\beta)\cos(\omega_{\rm d}t)$.
From~\cref{eq:C2_lambda_S5} we obtain
\begin{equation}
  \mathcal{C}_2(t,\beta)
  =2\,\lambda_0^2\,G(\beta)^2\int_0^t ds\;\cos^2(\omega_{\rm d}s)
  =2\,\lambda_0^2\,G(\beta)^2\left(\frac{t}{2}+\frac{\sin(2\omega_{\rm d}t)}{4\omega_{\rm d}}\right),
  \label{eq:C2_qubit_closed_S5}
\end{equation}
so that asymptotically $\mathcal{C}_2(t,\beta)\sim \lambda_0^2G(\beta)^2\,t$.
Combining~\cref{eq:C2_qubit_closed_S5} with the behavior of $\It$ yields an explicit gain-cost scaling.
In particular, on resonance $\omega_{\rm d}\to\Omega$ one has $\It\simeq m^2[\lambda_0G'(\beta)]^2t^2$, and thus
\begin{equation}
  \mathcal{R}_2(t,\beta)\;\sim\;
  m^2\left(\frac{G'(\beta)}{G(\beta)}\right)^{\!2}t
  \qquad (\omega_{\rm d}=\Omega,\ \lambda_0\ll\Omega).
  \label{eq:R2_resonant_scaling_S5}
\end{equation}
Away from resonance, the weak-field expansion yields a bounded, oscillatory $\It$ as a function of $t$,
while $\mathcal{C}_2(t,\beta)\sim t$; consequently $\mathcal{R}_2(t,\beta)$ is asymptotically suppressed as $1/t$ off resonance. This behaviour is shown explicitly in~\cref{fig:cost_tradeoff}, in which the information gain-cost ratio is plotted for different values of the spin-drive detuning.}

\teal{This shows how such a resource accounting provides a diagnostic tool for  characterizing when the information enhancement is ``energetically convenient" ($\mathcal{R}\geq1$)
At the same time, even in regimes where the protocol is not favorable according to this specific cost metric, the enhancement can still be operationally meaningful whenever increasing the absolute Fisher information relative to the equilibrium benchmark is the overriding objective (rather than minimizing control expenditure). This is especially relevant in temperature ranges where the equilibrium sensitivity of finite-dimensional probes is intrinsically small (e.g.\ at low temperatures), so that a controlled nonequilibrium boost can be valuable despite a higher control cost. In this sense, such a gain-cost proxy provides a transparent criterion for when the protocol is resource-favorable under control constraints, and when it should instead be viewed as a deliberate, resource-demanding route to improved thermometric sensitivity beyond equilibrium.}

\FloatBarrier

\end{document}